\documentclass[11pt,a4paper,twocolumn,english,prl,showpacs,preprintnumbers,amsmath,amssymb,nofootinbib]{revtex4}
\usepackage{lmodern}
\usepackage{lmodern}
\usepackage[T1]{fontenc}
\usepackage[latin9]{inputenc}
\usepackage{color}
\usepackage{babel}
\usepackage{verbatim}
\usepackage{amsmath}
\usepackage{amssymb}
\usepackage{esint}
\usepackage[unicode=true,pdfusetitle,
 bookmarks=true,bookmarksnumbered=false,bookmarksopen=false,
 breaklinks=false,pdfborder={0 0 1},backref=false,colorlinks=true]
 {hyperref}
\hypersetup{
 linkcolor=blue,citecolor=blue}

\makeatletter

\pdfpageheight\paperheight
\pdfpagewidth\paperwidth

\@ifundefined{textcolor}{}
{%
 \definecolor{BLACK}{gray}{0}
 \definecolor{WHITE}{gray}{1}
 \definecolor{RED}{rgb}{1,0,0}
 \definecolor{GREEN}{rgb}{0,1,0}
 \definecolor{BLUE}{rgb}{0,0,1}
 \definecolor{CYAN}{cmyk}{1,0,0,0}
 \definecolor{MAGENTA}{cmyk}{0,1,0,0}
 \definecolor{YELLOW}{cmyk}{0,0,1,0}
}

\usepackage{babel}
\usepackage{babel}
\usepackage{babel}

\@ifundefined{definecolor}{color}{}

\usepackage{latexsym}\usepackage{bm}

\makeatother

\begin{document}
\title{Non-Einsteinian Black Holes in Generic 3D Gravity Theories}
\author{Metin Gürses}
\email{gurses@fen.bilkent.edu.tr}

\affiliation{{\small{}Department of Mathematics, Faculty of Sciences}\\
{\small{}Bilkent University, 06800 Ankara, Turkey}}
\author{Tahsin Ça\u{g}r\i{} \c{S}i\c{s}man}
\email{tahsin.c.sisman@gmail.com}

\affiliation{Department of Astronautical Engineering,\\
 University of Turkish Aeronautical Association, 06790 Ankara, Turkey}
\author{Bayram Tekin}
\email{btekin@metu.edu.tr}

\affiliation{Department of Physics, Middle East Technical University, 06800 Ankara,
Turkey}
\begin{abstract}
The Bañados-Teitelboim-Zanelli (BTZ) black hole metric solves the
three-dimensional Einstein's theory with a negative cosmological constant
as well as all the generic higher derivative gravity theories based
on the metric; as such it is a universal solution. Here, we find,
in all generic higher derivative gravity theories, new universal non-Einsteinian
solutions obtained as Kerr-Schild type deformations of the BTZ black
hole. Among these, the deformed non-extremal BTZ black hole loses
its event horizon while the deformed extremal one remains intact as
a black hole in any generic gravity theory. 
\end{abstract}
\maketitle
\vspace{0.1cm}

\section{Introduction}

The black hole in 2+1 dimensions, the BTZ metric \citep{btz,BHTZ},
as a solution to vacuum Einstein's gravity with a negative cosmological
constant, shares many of the features of the $\left(3+1\right)$-dimensional
realistic Kerr black hole. Due to the local triviality of Einstein's
gravity in 2+1 dimensions, the BTZ solution has been a remarkable
tool in exploring the quantum nature of the black hole geometry such
as microscopic description of black hole entropy (see the review \citep{Carlip}
and the references therein). Three important features of the BTZ geometry
should be stressed. First, being a locally Einstein metric, it solves
all the metric based higher curvature gravity equations derived from
the most general action 
\begin{equation}
I=\int d^{3}x\sqrt{-g}\,{\cal L}\left(\text{Riem},\nabla\text{Riem},\cdots\right).\label{eq:Action}
\end{equation}
Such metrics are called universal which are unaffected by the quantum
effects \citep{GG1,GG2}. Generically, for dimensions greater than
three, Einstein metrics fail to solve higher derivative theories but
in three dimensions since the Riemann tensor can be written in terms
of the Einstein tensor $G_{\mu\nu}$ as $R_{\mu\alpha\nu\beta}=\epsilon_{\mu\alpha\sigma}\epsilon_{\nu\beta\sigma}G^{\sigma\rho}$,
any Einsteinian solution also solves the higher derivative theory
as long as the cosmological constant is tuned accordingly. This fact
is quite important and paves way to study the Einstein metrics such
as the BTZ black hole as solutions to the low energy quantum theory
of gravity at any scale defined by the action (\ref{eq:Action}) where
the nonmetric fields are set to zero or constant values. Secondly,
the BTZ geometry can be dressed with two arbitrary functions to represent
all the locally Einsteinian metrics yielding the Bañados geometry
as \citep{Banados} 
\begin{align}
ds^{2}=\ell^{2}\Biggl[\frac{{\rm d}r^{2}}{r^{2}}+ & \left(r{\rm d}u+\frac{1}{r}f\left(v\right){\rm d}v\right)\nonumber \\
 & \times\left(r{\rm d}v+\frac{1}{r}g\left(u\right){\rm d}u\right)\Biggr],\label{eq:Banados_geom}
\end{align}
where $u$ and $v$ are null coordinates. The geometry corresponds
to the non-extremal rotating BTZ black hole for constant nonvanishing
values of $f$ and $g$; and to the extremal rotating BTZ black hole
when one of these constants becomes zero. Thirdly, within the cosmological
Einstein's theory, the BTZ black hole has the uniqueness property
under the conditions described in \citep{Rooman,li}.

\noindent Due to the importance of the BTZ black hole, one would like
to know its uniqueness and also whether it is preserved as a black
hole under the deformations described as $g_{\mu\nu}=\bar{g}_{\mu\nu}+h_{\mu\nu}$
in the generic higher derivative theory (\ref{eq:Action}). Here,
$h_{\mu\nu}$ is not a small perturbation, hence just like the BTZ
black hole $\bar{g}_{\mu\nu}$, the deformed metric $g_{\mu\nu}$
is expected to solve the full field equations with the condition that
the black hole property is kept intact. Without a further specification
of the field equations of the theory, one cannot proceed further with
this most general deformation in a theory independent way. Therefore,
to keep the universal nature of the BTZ black hole under this deformation
in the setting of the most general higher derivative theory, we shall
consider a specific deformation which is called the Kerr-Schild--Kundt
(KSK) type whose universality; i.e. it solves the generic gravity
theory once a linear scalar partial differential equation is solved,
has been shown in \citep{Gurses-PRL,AdS-plane_pp-wave,KSK_universal}.
The KSK metric is in the form 
\begin{equation}
g_{\mu\nu}=\bar{g}_{\mu\nu}+2V\lambda_{\mu}\lambda_{\nu},\label{eq:AdS-waveKS}
\end{equation}
where $V$ is a scalar field and $\lambda$ is a null vector field
which satisfy the properties 
\begin{align}
\lambda^{\mu}\lambda_{\mu} & =0,\quad\nabla_{\mu}\lambda_{\nu}\equiv\xi_{(\mu}\lambda_{\nu)},\nonumber \\
\xi_{\mu}\lambda^{\mu} & =0,\quad\lambda^{\mu}\partial_{\mu}V=0,\label{eq:AdS-wave_prop}
\end{align}
for both the background and the full metric. The $\xi$ vector is
defined via the second equation in (\ref{eq:AdS-wave_prop}) once
the $\lambda$ null vector is chosen (a way to generate viable $\lambda$
vectors from smooth curves was given in \citep{smooth}). For the
KSK metrics, the Ricci tensor becomes 
\[
R_{\mu\nu}=\left(\mathcal{Q}V\right)\lambda_{\mu}\lambda_{\nu}-\frac{2}{\ell^{2}}g_{\mu\nu},
\]
where $\ell$ is the AdS length and the operator $\mathcal{Q}$ is
defined as 
\[
\mathcal{Q}V\equiv\left(\bar{g}^{\mu\nu}\bar{\nabla}_{\mu}\bar{\nabla}_{\nu}+2\xi^{\mu}\partial_{\mu}+\frac{1}{2}\xi^{\mu}\xi_{\mu}-\frac{2}{\ell^{2}}\right)V.
\]
Then, for the pure cosmological Einstein theory, the nonlinear field
equations $R_{\mu\nu}=2\Lambda g_{\mu\nu}$ become linear in $V$
and boil down to \citep{gurses1} 
\begin{equation}
\mathcal{Q}V=0,\label{eq:Einstein_operator}
\end{equation}
once the trace of the field equationsis is solved as $\Lambda=-1/\ell^{2}$.
Given the background metric in some local coordinates, one can find
the local solution. For a general gravity theory with the highest
derivative order of $\left(2N+2\right)$ in the field equations with
$N\ge0$, the field equations reduce to \citep{Gurses-PRL,AdS-plane_pp-wave,GravWaves3D}
\begin{equation}
\prod_{n=1}^{N}\,\big({\cal Q}-m_{n}^{2}\big)\,{\cal Q}\,V=0,\label{denk2}
\end{equation}
whose generic solution is $V=V_{E}+\sum_{n=1}^{N}\,V_{n}$ where the
Einsteinian part ($V_{E}$) and the other (massive) parts, assuming
nondegeneracy, satisfy the following equations 
\begin{eqnarray}
{\cal Q}V_{E}=0,\qquad\big({\cal Q}-m_{n}^{2}\big)\,V_{n}=0.\label{denk3}
\end{eqnarray}
One can also interpret these equations as transverse-traceless perturbations
of the background space, therefore they correspond to massless and
massive gravitons. In three dimensional Einstein's theory, since there
are no gravitons, $V_{E}$ corresponds to pure gauge transformations
when the deformation $h_{\mu\nu}$ is assumed to be a perturbation
about the exact background. On the other hand, the $V_{n}$ solutions
are the non-Einsteinian solutions with the Ricci tensor $R_{\mu\nu}=\left(\sum_{n=1}^{N}m_{n}^{2}V_{n}\right)\lambda_{\mu}\lambda_{\nu}-2/\ell^{2}g_{\mu\nu}$.

\vspace{0.1cm}

\section{Deformations of BTZ}

Along the lines described above, let us consider the deformations
of the BTZ black hole 
\begin{equation}
d\bar{s}^{2}=-h{\rm d}t^{2}+\frac{{\rm d}r^{2}}{h}+r^{2}\left({\rm d}\phi-\frac{j}{2r^{2}}{\rm d}t\right)^{2},\label{btz1}
\end{equation}
with $h\left(r\right)=-m+\frac{r^{2}}{\ell^{2}}+\frac{j^{2}}{4r^{2}}$.
We shall call the generic deformation as BTZ-waves since the general
solution will be of the wave form depending on the null coordinates.
As we shall show below, among these only a subclass will remain a
black hole. In (\ref{btz1}), $m$ and $j$ are constants representing
the mass and angular momentum, respectively. The outer and inner horizons
of the black hole are located at 
\begin{equation}
r_{\pm}^{2}=\frac{m\ell^{2}}{2}\left(1\pm\sqrt{1-\frac{j^{2}}{m^{2}\ell^{2}}}\right).
\end{equation}
which coalesce for the extremal case $j=\pm m\ell$ at $r_{0}^{2}=m\ell^{2}/2$.

\noindent To understand if and how the black hole nature of the BTZ
metric is changed by the KSK deformation, let us study the event horizon.
In the generic case, the symmetries of the BTZ geometry are no longer
symmetries of the KSK geometry. Hence, the detection of the event
horizon cannot be done with the Killing vectors; instead, since the
horizons will be null hypersurfaces defined as level sets of $r$,
let us consider where the surface normal $\partial_{\mu}r$ becomes
a null vector in the BTZ-wave geometry as

\begin{equation}
\Omega\equiv g^{\mu\nu}\partial_{\mu}r\partial_{\nu}r=0.
\end{equation}
Using (\ref{eq:AdS-waveKS}) and (\ref{btz1}), $\Omega$ becomes

\begin{align}
\Omega & =h\left(r\right)-2V\left(\lambda^{\mu}\partial_{\mu}r\right)^{2}\nonumber \\
 & =2V\left(t,r_{\pm},\phi\right)\left(\left.\lambda^{r}\right|_{r=r_{\pm}}\right)^{2}.
\end{align}
Here, to have $\Omega=0$, $V\left(t,r_{\pm},\phi\right)=0$ is a
possibility but recall that the metric function $V$ must satisfy
a theory dependent differential equation. Then, to keep the BTZ black
hole intact in a theory independent way, 
\begin{equation}
\left.\lambda^{r}\right|_{r=r_{\pm}}=0,\label{eq:BH_cond}
\end{equation}
must be satisfied. In this way, one has the deformed black hole solutions
for all generic gravity theories. There can be other black hole solutions
where the profile function $V$ satisfies the condition $V\left(t,r_{\pm},\phi\right)=0$
or $h\left(r\right)-2V\left(\lambda^{\mu}\partial_{\mu}r\right)^{2}=0$
for different $r$ values. In these cases, since $V$ takes different
functional forms in different gravity theories then such black holes
solutions will be theory dependent; and given the theory, one can
construct these.

Since we are interested in the KSK-type deformations of the BTZ black
hole which keep the event horizon intact, we considered a null hypersurface
of constant $r$ to locate the event horizon. However, for the KSK
metric (\ref{eq:AdS-waveKS}), in general, a null hypersurface of
the form $F\left(t,r,\theta\right)={\rm constant}$ should be considered
to locate a horizon as 
\begin{equation}
\bar{g}^{\mu\nu}\partial_{\mu}F\partial_{\nu}F-2V\left(t,r,\theta\right)\left(\lambda^{\mu}\partial_{\mu}F\right)^{2}=0.\label{eq:General_horizon}
\end{equation}
In addition to the undeformed event horizon given with (\ref{eq:BH_cond}),
to have a deformed horizon for the KSK geometry, this equation should
be considered which will be studied elsewhere \citep{BTZwaves}. The
analysis of (\ref{eq:General_horizon}) in its full generality is
a tedious task; however, to get some understanding, for the $\lambda_{\mu}=\partial_{\mu}u$
case,\footnote{This choice is motivated at the beginning of the next section.}
let us consider the $r=f\left(u\right)$ hypersurface which becomes
null if%
\begin{equation}
0=h\left(r\right)-2\lambda^{r}\left(\frac{df}{du}+V\left(t,r,\theta\right)\lambda^{r}\right).\label{eq:r=00003Df(u)_hypersurf}
\end{equation}
To have an equation in $r$ and $u$ with a solution $r=f\left(u\right)$,
one must have $V=V\left(u,r\right)$ with a $\lambda^{r}$ depending
only $r$. Then, the KSK property $\lambda^{\mu}\partial_{\mu}V=0$
reduces to%
\begin{equation}
\lambda^{r}\frac{\partial V}{\partial r}=0,
\end{equation}
which requires either $\lambda^{r}=0$ or $V=V\left(u\right)$. For
$\lambda^{r}=0$, (\ref{eq:r=00003Df(u)_hypersurf}) becomes $h\left(r\right)=0$
so it does not provide a generalization as $r=f\left(u\right)$. Thus,
one needs to have $V=V\left(u\right)$ in general. For this case,
$\partial_{\mu}V=V_{u}\lambda_{\mu}$ and $\bar{g}^{\mu\nu}\bar{\nabla}_{\mu}\bar{\nabla}_{\nu}V=0$,
so $\mathcal{Q}V$ reduces
\begin{equation}
\mathcal{Q}V=\left(\frac{1}{2}\xi^{\mu}\xi_{\mu}-\frac{2}{\ell^{2}}\right)V.
\end{equation}
To obtain an Einsteinian solution, $\mathcal{Q}V=0$ must be satisfied
which is the case for any $V=V\left(u\right)$ if $\xi^{2}=4/\ell^{2}$.
The condition $\xi^{2}=4/\ell^{2}$ is satisfied for the BTZ-waves
constructed in the next section. Thus, one may find a solution for
(\ref{eq:r=00003Df(u)_hypersurf}) indicating a null hypersurface
of the form $r=f\left(u\right)$ exists if $V=V\left(u\right)$. However,
the $V=V\left(u\right)$ solution is an Einsteinian metric which is
already represented in the Bañados geometry. Note that this case also
covers the shifted horizons, that is $r={\rm constant}$ but $r\ne r_{\pm}$,
by having $f={\rm constant}$ and $V={\rm constant}$. On the other
hand, to have a non-Einsteinian KSK geometry for an horizon of the
form $r=f\left(u\right)$, the metric function $V=V\left(u\right)$
must satisfy 
\begin{equation}
\big({\cal Q}-m_{n}^{2}\big)\,V_{n}=\left(\frac{1}{2}\xi^{\mu}\xi_{\mu}-\frac{2}{\ell^{2}}-m_{n}^{2}\right)V=0,\label{eq:Non-Einsteinian_eqn_for_V(u)}
\end{equation}
where $m_{n}$ depends on the parameters of the higher derivative
theory. Since $\xi^{\mu}$ is theory independent, (\ref{eq:Non-Einsteinian_eqn_for_V(u)})
cannot be satisfied in general. Therefore, it is not possible to obtain
a non-Einsteinian KSK geometry that has a horizon of the form $r=f\left(u\right)$.
In \citep{BTZwaves}, we will study more general horizon forms such
as $r=f\left(u,\psi\right)$ with $\xi_{\mu}=\partial_{\mu}\psi$
which require more general $V$ beyond $V=V\left(u\right)$.

In the discussion below, we will show that the condition (\ref{eq:BH_cond}),
which keeps the BTZ event horizon intact, can be satisfied if and
only if the BTZ seed is extremal so that a subclass of the BTZ-waves
will be a deformed version of the extremal BTZ black hole.

\vspace{0.1cm}

\section{BTZ-wave construction}

Now, let us obtain the BTZ-wave metrics by a direct construction.
As a consequence of the second property in (\ref{eq:AdS-wave_prop}),
let us choose the null one-form field $\lambda_{\mu}$ to be exact,
$\lambda_{\mu}=\partial_{\mu}u\left(t,r,\phi\right)$. Then, the condition
that $\lambda_{\mu}$ be null yields 
\begin{align}
-\left(\frac{\partial u}{\partial t}\right)^{2}-\frac{j}{r^{2}}\frac{\partial u}{\partial t}\frac{\partial u}{\partial\phi}\label{eq:Nullity}\\
+\left(\frac{h}{r^{2}}-\frac{j^{2}}{4r^{4}}\right)\left(\frac{\partial u}{\partial\phi}\right)^{2}+h^{2}\left(\frac{\partial u}{\partial r}\right)^{2} & =0.\nonumber 
\end{align}
Notice that all coefficients are a function of $r$, so the easiest
way to satisfy the nullity condition is to consider a $u$ whose derivatives
are either a function of $r$ or a constant as \footnote{There can be other choices for the function $u$ providing different
solutions which will be discussed elsewhere \citep{BTZwaves}.} 
\begin{equation}
u\left(t,r,\phi\right)=c_{1}t+c_{2}\phi+w\left(r\right).\label{eq:ansatz}
\end{equation}
This ansatz provides a solvable set of differential equations for
the KSK metric properties. The solution can be put in a simpler form
if the BTZ metric is written in terms of $r_{\pm}$ with $h\left(r\right)=\frac{\left(r^{2}-r_{+}^{2}\right)\left(r^{2}-r_{-}^{2}\right)}{r^{2}\ell^{2}}$
and $j=\frac{2\sigma r_{+}r_{-}}{\ell}$ where $\sigma$ represents
the direction of rotation which we choose to be $\sigma=+1$. For
this non-extremal BTZ seed, the $\lambda_{\mu}$ and $\xi_{\mu}$
one-forms are found to be 
\begin{equation}
\lambda_{\mu}=\left(1,\frac{\ell^{2}r\left(r_{+}+\epsilon r_{-}\right)}{\left(r^{2}-r_{+}^{2}\right)\left(r^{2}-r_{-}^{2}\right)},\epsilon\ell\right),\label{eq:lambda_nonextremal}
\end{equation}
and 
\[
\xi_{\mu}=\left(-\frac{r_{+}+\epsilon r_{-}}{\ell^{2}},-\frac{r\left(\alpha+\beta\right)}{\ell^{2}\alpha\beta},\frac{\epsilon r_{+}+r_{-}}{\ell}\right),
\]
where $\epsilon$ is equal to $\pm1$, $\alpha$ and $\beta$ are
defined as $\alpha\left(r\right)=\left(r^{2}-r_{+}^{2}\right)/\ell^{2}$
and $\beta\left(r\right)=\left(r^{2}-r_{-}^{2}\right)/\ell^{2}$.
From (\ref{eq:lambda_nonextremal}), $\lambda^{r}$ can be calculated
to be 
\begin{equation}
\lambda^{r}=h\left(r\right)\lambda_{r}=\frac{r_{+}+\epsilon r_{-}}{r}.
\end{equation}
The black hole event horizon condition (\ref{eq:BH_cond}) is not
satisfied, so the BTZ deformation for the nonextremal case is not
a black hole in the generic theory. Yet, the resulting metric is a
solution to the generic theory if $V$ satisfies the constraint $\lambda^{\mu}\,\partial_{\mu}V=0$
and (\ref{denk2}) for the specific theory. The constraint can be
solved in a theory independent way and the solution is 
\begin{align}
V\left(t,r,\phi\right)=\mathcal{F}\Biggl( & t+\frac{r_{+}\ln\alpha-\epsilon r_{-}\ln\beta}{2\left(\beta-\alpha\right)},\nonumber \\
 & \phi+\frac{r_{-}\ln\alpha-\epsilon r_{+}\ln\beta}{2\left(\beta-\alpha\right)}\Biggr),\label{eq:Mthd_char_soln}
\end{align}
where $\mathcal{F}$ is a smooth function.

\noindent Above, we discussed the nonextremal case, now let us focus
to the extremal case $j=m\ell$ with $h\left(r\right)=\frac{\left(r^{2}-r_{0}^{2}\right)^{2}}{\ell^{2}r^{2}}$
and $j=\frac{2r_{0}^{2}}{\ell}$. For this case, the sign choice $\epsilon$
becomes important as one arrives at two different metrics. For $\epsilon=+1$,
with a similar construction as in the nonextremal case, the $\lambda_{\mu}$
and $\xi_{\mu}$ one-forms become 
\begin{equation}
\lambda_{\mu}=\left(1,\frac{2rr_{0}\ell^{2}}{\left(r^{2}-r_{0}^{2}\right)^{2}},\ell\right),\label{eq:Extremal_lambda_nonzero_r_comp}
\end{equation}
and 
\[
\xi_{\mu}=\left(-\frac{2r_{0}}{\ell^{2}},-\frac{2r}{r^{2}-r_{0}^{2}},\frac{2r_{0}}{\ell}\right).
\]
From (\ref{eq:Extremal_lambda_nonzero_r_comp}), $\lambda^{r}$ can
be calculated to be 
\begin{equation}
\lambda^{r}=\frac{2r_{0}}{r}.
\end{equation}
Again, the black hole event horizon condition (\ref{eq:BH_cond})
is not satisfied, so the BTZ deformation for the extremal case with
$\epsilon=+1$ is not a black hole in the generic theory.

For $\epsilon=-1$, the KSK metric construction for the extremal case
differs in a subtle way from the nonextremal construction such that
(\ref{eq:Nullity}) requires $w\left(r\right)$ in (\ref{eq:ansatz})
to be constant. As a result, the $\lambda_{\mu}$ and $\xi_{\mu}$
one-forms become 
\begin{equation}
\lambda_{\mu}=\left(1,0,-\ell\right),\label{eq:Extremal_lambda_zero_r_comp}
\end{equation}
and 
\begin{equation}
\xi_{\mu}=\left(0,\frac{2r}{r_{0}^{2}-r^{2}},0\right).\label{eq:ksi}
\end{equation}
From (\ref{eq:Extremal_lambda_zero_r_comp}), $\lambda^{r}$ can simply
be found to be 
\begin{equation}
\lambda^{r}=0.
\end{equation}
This time, the black hole event horizon condition (\ref{eq:BH_cond})
is satisfied, so the BTZ deformation for the extremal case with $\epsilon=-1$
is a black hole in the generic theory. Here, the metric function $V$
must satisfy $\lambda^{\mu}\partial_{\mu}V=0$ yielding 
\begin{equation}
\frac{\ell^{2}}{r^{2}-r_{0}^{2}}\left(\ell\frac{\partial V}{\partial t}+\frac{\partial V}{\partial\phi}\right)=0.
\end{equation}
with the solution 
\begin{equation}
V=V\left(t-\ell\phi,r\right).
\end{equation}
The explicit form of $V$ will be given below for Einstein's theory
and the new massive gravity (NMG) \citep{NMG}.

\vspace{0.1cm}

\subsection{Extremal-BTZ Wave Solution of Einstein's Gravity}

We showed that the only possible KSK deformation of BTZ black hole
which keeps the black hole nature intact is the extremal BTZ black
hole deformed with the constant null vector field of $\lambda_{\mu}=\left(1,0,-\ell\right)$.
Now, let us find the metric function $V$ for the cosmological Einstein's
gravity by solving (\ref{eq:Einstein_operator}). With (\ref{eq:ksi}),
the field equation for $V$ becomes 
\begin{equation}
r\frac{\partial^{2}}{\partial r^{2}}V_{E}\left(u,r\right)-\frac{\partial}{\partial r}V_{E}\left(u,r\right)=0,
\end{equation}
where we defined $u=t-\ell\phi$ which is in fact the generating function
for $\lambda_{\mu}$ as $\lambda_{\mu}=\partial_{\mu}u$. If $r\ne r_{0}$,
the Einsteinian solution becomes 
\begin{equation}
V_{E}\left(u,r\right)=c_{1}\left(u\right)r^{2}+c_{2}\left(u\right),\label{eq:V_E}
\end{equation}
yielding the metric 
\begin{align*}
ds^{2}= & d\bar{s}^{2}+2\left(c_{1}\left(u\right)r^{2}+c_{2}\left(u\right)\right)\left({\rm d}t-\ell\,{\rm d}\phi\right)^{2},
\end{align*}
where $d\bar{s}^{2}$ is the extremal BTZ seed. This result is consistent
with the Bañados geometry (\ref{eq:Banados_geom}) and the analysis
of \citep{li}. As in the case of the Bañados geometry which dresses
the BTZ black hole with two arbitrary functions, our generic solution
with arbitrary $c_{1}\left(u\right)$ and $c_{2}\left(u\right)$ are
of the nonlinear wave type which we called the BTZ wave. To understand
this solution better, we can compute its mass and angular momentum
using the Abbott-Deser approach \citep{Abbott}. Assuming $c_{1}\left(t-\ell\phi\right)=c_{2}\left(t-\ell\phi\right)=0$
and $r_{0}=0$ to be the background, the mass corresponding to the
background time-like Killing vector $\zeta^{\mu}=\left(-1,0,0\right)$
is $M=m+\frac{2}{\pi}\int_{0}^{2\pi}d\phi\,c_{2}\left(t-\ell\phi\right)$;
and the angular momentum corresponding to the background Killing vector
$\zeta^{\mu}=\left(0,0,1\right)$ is $J=m\ell+\frac{2\ell}{\pi}\int_{0}^{2\pi}d\phi\,c_{2}\left(t-\ell\phi\right)$.
We have kept mass and angular momentum computation with generic $c_{1}\left(u\right)$
and $c_{2}\left(u\right)$. Since this solution is no longer stationary,
its mass angular momentum are time dependent via these functions.
Note that the extremality condition is intact as $J=M\ell$. The function
$c_{1}\left(t-\ell\phi\right)$ corresponds to a pure gauge and does
not appear in the mass and angular momentum expressions. Of course,
for a stationary black hole solution, the arbitrary $u$ dependent
functions should be taken as constants as we mentioned for (\ref{eq:Banados_geom}).
Then, one obtains time-independent mass and angular momentum. The
discussion is exactly like the case of Bañados metric \citep{Banados,li}.

\noindent \vspace{0.1cm}

\subsection{Extremal-BTZ Wave Solution of}

\emph{NMG} Now, we study the solution of cosmological new massive
gravity (NMG) given with the action 
\begin{equation}
I=-\frac{1}{\kappa^{2}}\int d^{3}x\,\sqrt{-g}\left(R-2\Lambda_{0}+L^{2}K\right),
\end{equation}
whose field equations are 
\begin{equation}
G_{\mu\nu}+\Lambda_{0}g_{\mu\nu}-\frac{L^{2}}{2}K_{\mu\nu}=0,\label{eq:NMG_eom}
\end{equation}
where $K_{\mu\nu}=2\square R_{\mu\nu}-\frac{1}{2}\left(\nabla_{\mu}\nabla_{\mu}+g_{\mu\nu}\square\right)R+4R_{\mu\alpha\nu\beta}R^{\alpha\beta}-\frac{3}{2}RR_{\mu\nu}-g_{\mu\nu}K$
and the trace $K=g^{\mu\nu}K_{\mu\nu}=R_{\mu\nu}R^{\mu\nu}-\frac{3}{8}R^{2}$.
Putting the metric of the extremal BTZ wave defined by $\lambda_{\mu}$
given in (\ref{eq:Extremal_lambda_zero_r_comp}) yields the field
equations 
\begin{align}
\frac{1}{\ell^{2}}+\Lambda_{0}+\frac{L^{2}}{4\ell^{4}} & =0,\label{eq:Tr_eqn}\\
\big({\cal Q}-m_{g}^{2}\big){\cal Q}V & =0,\label{eq:Trless_eqn}
\end{align}
where $m_{g}^{2}$ is the mass of the spin-2 graviton of the NMG theory
given as 
\begin{equation}
m_{g}^{2}=\frac{1}{L^{2}}-\frac{1}{2\ell^{2}}.
\end{equation}
The first equation determines the effective cosmological parameter
$\ell$. The second equation (\ref{eq:Trless_eqn}) determines the
metric function $V$ and has the general solution 
\begin{equation}
V\left(u,r\right)=V_{E}\left(u,r\right)+V_{p}\left(u,r\right),\label{eq:General_soln}
\end{equation}
where $u=t-\ell\phi$ and $V_{E}$ is the Einsteinian solution (\ref{eq:V_E})
while $V_{p}$ is the solution of the massive operator $\big({\cal Q}-m_{g}^{2}\big)$
which can be found as 
\begin{align}
V_{p}\left(u,r\right)= & c_{3}\left(u\right)\left(r^{2}-r_{0}^{2}\right)^{\left(1+p\right)/2}\nonumber \\
 & +c_{4}\left(u\right)\left(r^{2}-r_{0}^{2}\right)^{\left(1-p\right)/2},
\end{align}
with $p\equiv\sqrt{m_{g}^{2}\ell^{2}+1}$. The reality of $p$ is
equivalent to the Breitenlohner-Freedman (BF) bound \citep{BF}. It
is important to note that the solution (\ref{eq:General_soln}) to
this quadratic theory solves all higher curvature theories as long
as the corresponding effective cosmological constant equation is satisfied.
Using the %
construction of \citep{Deser_Tekin-PRL}, one can show that the finiteness
of mass and angular momentum requires $c_{3}\left(u\right)=c_{4}\left(u\right)=0$
for $0<p<1$, $c_{3}\left(u\right)=0$ for $1<p$, or $c_{4}\left(u\right)=0$
for $p<-1$ yielding the mass $M=m\left(1+\frac{2}{2p^{2}-1}\right)$
and the angular momentum $J=M\ell$ such that extremality is kept
intact.

\noindent \vspace{0.1cm}

\section{Conclusions}

We have studied the exact deformation of the BTZ black hole in the
context of generic gravity; and showed that the non-extremal black
hole loses its exact horizon and the resulting deformed metric is
of wave type, which we called the BTZ wave. Surprisingly, the deformed
extremal black hole remains a black hole. There are several ways to
read this result: First, the non-extremal BTZ is unique in generic
gravity while the extremal one is not as in the case of Einstein's
theory; secondly, considering the deformations as generic quantum
or classical corrections, the non-extremal BTZ is not preserved as
a black hole solution to the generic gravity while the extremal one
remains a black hole in any generic gravity theory. Lastly, regarding
the $r=0$ singularity after the KSK deformation, note that all the
curvature invariants of the KSK metrics are constant; therefore, there
is no curvature singularity.

\end{document}